\documentclass[]{aa}
\usepackage{graphicx}
\usepackage{subfigure}
\usepackage{amsmath}
\usepackage{amssymb}
\def\kms{\relax \ifmmode {\,\mbox{km\,s}}^{-1}\else \,\mbox{km\,s}$^{-1}$\fi}
\def\ha{\relax \ifmmode {\mbox H}\alpha\else H$\alpha$\fi}
\def\hb{\relax \ifmmode {\mbox H}\beta\else H$\beta$\fi}
\def\hi{\relax \ifmmode {\mbox H\,{\scshape i}}\else H\,{\scshape i}\fi}  
\def\hii{\relax \ifmmode {\mbox H\,{\scshape ii}}\else H\,{\scshape ii}\fi}
\def\oiii{\relax \ifmmode {\mbox O\,{\scshape iii}}\else O\,{\scshape iii}\fi}
\def\oii{\relax \ifmmode {\mbox O\,{\scshape ii}}\else O\,{\scshape ii}\fi}
\def\oi{\relax \ifmmode {\mbox O\,{\scshape i}}\else O\,{\scshape i}\fi}
\def\nii{\relax \ifmmode {\mbox N\,{\scshape ii}}\else N\,{\scshape ii}\fi}
\def\sii{\relax \ifmmode {\mbox S\,{\scshape ii}}\else S\,{\scshape ii}\fi}
\def\lha{\relax \ifmmode \mbox {L}_{H\alpha}\else $\mbox{L}_{H\alpha}$\fi}
\def\ldig{\relax \ifmmode {\mbox L}_{DIG}\else ${\mbox L}_{DIG}$\fi}
\def\ls{\relax \ifmmode {\mbox L}_{ Str}\else ${\mbox L}_{ Str}$\fi}
\def\eme{\relax \ifmmode {\,\mbox{pc\,cm}}^{-6}\else \,pc\,cm$^{-6}$\fi}
\def\l{\relax \ifmmode  \lambda\else $\lambda$\fi}

\def\Msun{M$_\odot$}

\def\me{$^{-1}$}              
\def\arcmin{\hbox{$^\prime$}}
\def\arcsec{\hbox{$^{\prime\prime}$}}
\def\deg{\hbox{$^\circ$}}
\def\fs{\hbox{$^{\rm s}$}}
\def\hms#1h#2m#3s{\relax \ifmmode #1^{\rm h}\,#2^{\rm m}\,#3^{\rm s}
                   \else \hbox{$#1^{\rm h}\,#2^{\rm m}\,#3^{\rm s}$}
                  \fi}
\def\dms#1d#2m#3s{\relax#1\deg\,#2\arcmin\,#3\arcsec}
\def\hmsd#1h#2m#3.#4s{\relax\ifmmode #1^{\rm h}\,#2^{\rm m}\,#3.#4\fs
                      \else \hbox{$#1^{\rm h}\,#2^{\rm m}\,#3#4\fs$}
                      \fi}
\begin{document}
   \title{Ionized gas kinematics and massive star formation in NGC~1530}

  \author{A. Zurita\inst{1}, M. Rela\~no\inst{2}, J. E. Beckman\inst{2,3}, and J. H. Knapen\inst{4}}   
   \offprints{J.E. Beckman (jeb@ll.iac.es)}

   \institute{Isaac Newton Group of Telescopes, 38700, Santa Cruz de La Palma, Canarias, Spain\\
       \email{azurita@ing.iac.es}
     \and Instituto de Astrof\'\i sica de Canarias, C. V\'\i a L\'actea s/n,
       38200, La Laguna, Tenerife, Spain \\
       \email{mpastor@ll.iac.es, jeb@ll.iac.es}
     \and Consejo Superior de Investigaciones Cient\'\i ficas (CSIC), Spain 
     \and University of Hertfordshire,
         Hatfield, Herts. AL10 9AB, UK \\
       \email{knapen@star.herts.ac.uk}}
   \date{}
\abstract{
We present emission line mapping of the strongly barred galaxy NGC~1530 
obtained using Fabry--P\'erot interferometry in \ha, at significantly enhanced
angular resolution compared with  previously published studies. The main point of 
the work is to examine in detail the non--circular components of the velocity 
field of the gas, presumably induced by the strongly non--axisymmetric gravitational 
potential of the bar. To do this we first derive a model rotation curve making
minimum assumptions about kinematic symmetry, and go on to measure the 
non--circular component of the full radial velocity field. This clearly reveals
the streaming motions associated with the spiral density wave producing the
arms, and the quasi--elliptical motions with speeds of order 100 km s\me\
aligned with the bar. It also shows in some detail how these flows swing in
towards and around the nucleus as they cross a circumnuclear resonance, from the 
dominant ``$x_1$ orbits" outside the resonance to ``$x_2$ orbits" within it. Comparing
cross--sections of this residual velocity map along and across the bar with
the surface brightness map in \ha\ indicates a systematic offset between
regions of high non--circular velocity and massive star formation. To 
investigate further we produce maps of velocity gradient along and across the
bar. These illustrate very nicely the shear compression of the gas, revealed by
the location of the dust lanes along loci of maximum velocity gradient 
perpendicular to the bar. They also show clearly how shear, seen in our data 
as velocity gradient perpendicular to the flow, acts to inhibit massive star
formation, whereas shocks, seen as strong velocity gradients along the flow
vector, act to enhance it. Although the inhibiting effect of gas shear flow on
star formation has long been predicted, this is the clearest observational
illustration so far of the effect, thanks to the strong shock--induced
counterflow system in the bar. It is also the clearest evidence that dust picks
out shock--induced inflow along bars. These observations should be of 
considerable interest to those modelling massive star formation in general.
}
\date{} \authorrunning{Zurita et al.}  
\maketitle
%
%________________________________________________________________
\section{Introduction}
\label{intro}

The interactive relation between dynamics and morpho\-lo\-gy is a well known 
aspect of barred galaxies. It has been invoked to throw light on one of the 
more interesting questions in galactic evolution: by what mechanism or
mechanisms is material driven in to regions close to the nucleus where 
it can participate in star formation, or contribute to the mass of
the central supermassive black hole, and to the activity around it
(Shlosman et al. 1989). As an archetypal barred galaxy NGC~1530 has 
received con\-si\-de\-rable attention both morphologically 
and kinematically. These stu\-dies have been facilitated observationally by the strength and 
length of the main bar, which compensate for its distance. The latter is
36.6~Mpc (Tully \& Fisher 1988), which agrees with a more recent calculation of
37~Mpc using its recession velocity of 2666~\kms (Young et al. 1989) and a 
value for H$_{\rm o}$ of 72~\kms~Mpc$^{-1}$ (Freedman et al. 2001). In spite of this, the bar and 
its accompanying kinematics can be well observed because its deprojected radial 
length is some 14~kpc. The
strength of the bar (it has one of the highest bar torques measured by Buta
et al. (2003)) has attracted the attention of observers to the distinctive
phenomenology of this galaxy. 
In recent years two groups have made major contributions to the study of NGC~1530 
which are particularly relevant to the work presented here. 
They are Regan et al. (1995, 1996, 1997) and the group of Downes \& Reynaud 
(Downes et al. 1996; Reynaud \& Downes 1997, 1998, 1999; Greve et al. 1999). 
The types of observations these groups present overlap to some degree.
The group of Downes \& Reynaud base most of their work on velocity and
surface luminosity mapping in CO with some additional stellar information 
presented in Greve et al. (1999), while Regan et al.'s most important
contributions come from their velocity and surface brightness mapping in \ha\ 
with a useful contribution from CO supplemented by near--IR mapping in the
initial paper of their published series.

\begin{figure*}
\centering
\vspace{-0.5cm}
\includegraphics[width=\textwidth]{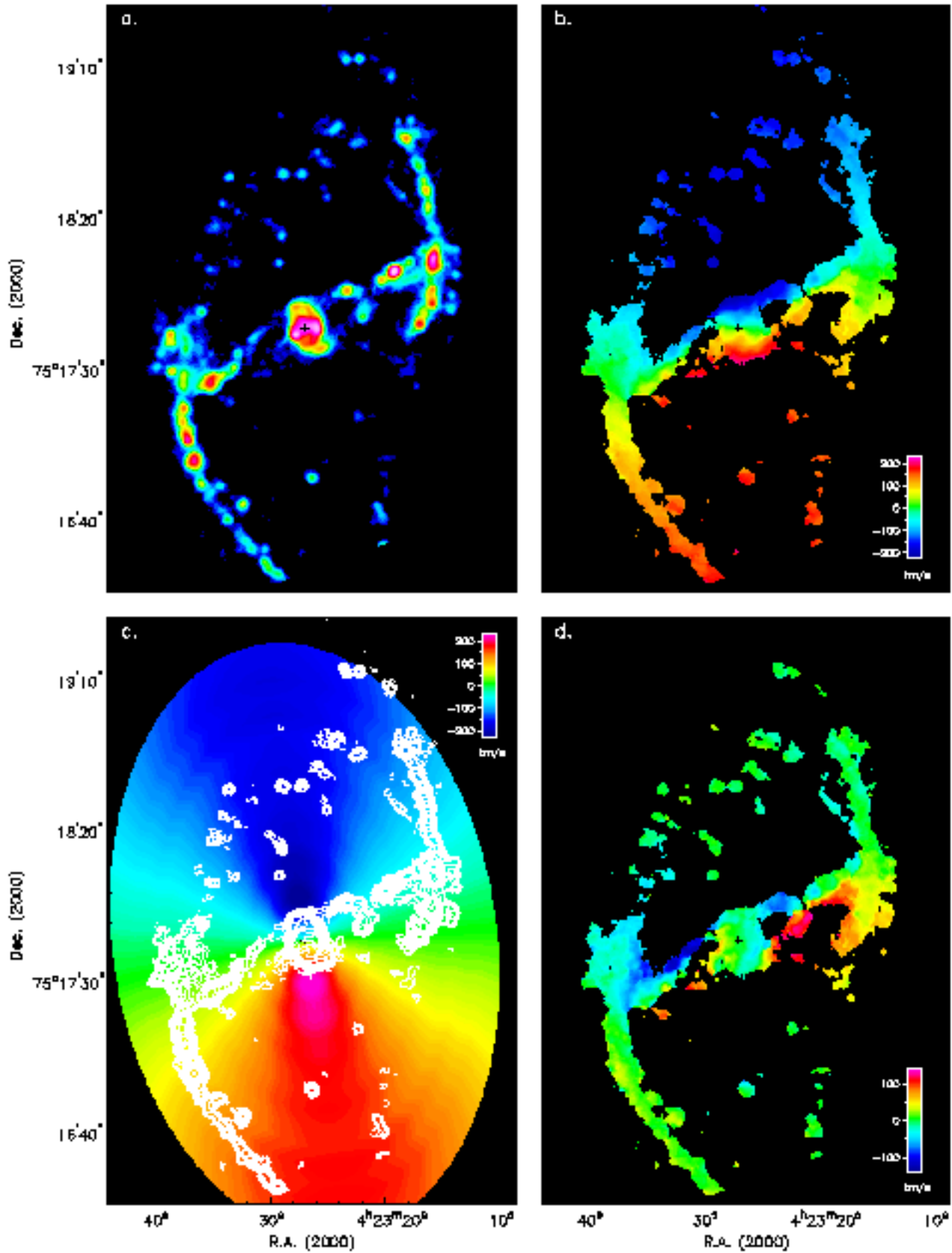}
\vspace{-1cm}
\caption{ {\bf a)} Intensity map (zeroth order moment) of the \ha\ emission in NGC~1530. 
The map was obtained from the 2\arcsec\ resolution data cube. The kinematic
centre is marked with a black cross. {\bf b)} Velocity map (first moment) of
the \ha\ emission in NGC~1530, derived from the 2\arcsec\ resolution data
cube.  {\bf c)} Model
velocity map of NGC~1530, obtained from the \ha\ rotation curve of the galaxy 
shown in Fig.~\ref{curva}, with \ha\ isointensity contours overlaid. 
We can clearly see ``wiggles'' at the radii of the 
spiral arms, due to the density wave streaming motions not subtracted off 
from the rotation curve. {\bf d)} Map of the residual velocities
of NGC~1530, obtained by subtracting the model velocity field  (Fig.~\ref{mosaicmaps}c, after taking 
out the ripples due to the streaming motions in the spiral arms, see text for details)
from the velocity map (Fig.~\ref{mosaicmaps}b).}
\label{mosaicmaps}
\end{figure*}

NGC~1530 is a dramatic twin-armed spiral galaxy, with arms breaking away 
sharply from the ends of the long primary bar. It is notable that these arms,
though\- trailing, ha\-ve shorter opposing lengths, curving inward from the 
tips of the bar as prolongations of the main arms. They are well brought out
in \ha\ as seen in Regan et al. (1995) or in Fig.~\ref{mosaicmaps}a of the present
paper. The bar ima\-ged in broad band is wide and contains prominent dust lanes 
which, as the publications by Regan et al. and by Downes \& Reynaud listed above 
well point out, pick out shocked gas following trajectories generally aligned
with the bar. These are almost straight along the bar but begin to curve
in\-wards from a galactocentric distance of some 2~kpc,
bending sharply in towards the nucleus. They link the prominent star--forming 
regions at the ends of the bar, seen very well here in \ha\ in
Fig.~\ref{mosaicmaps}a, with the even more prominent star--forming ring feature 
with radius $\sim$2~kpc noted in blue images by Buta (1984) and by Wray
(1988) and also clearly picked out in \ha\ in Fig.~\ref{mosaicmaps}a. Combining
near--IR with optical data to give colour maps, Regan et al. (1995) explored
the stellar and dust morphology in the bar and in the central zone following 
the lanes right into the nucleus. One of the aims of the present paper is to 
elucidate further the relation between gas flows and star formation which are 
implied in these morphological data described above.

However, this aim has been pursued with con\-si\-de\-rable achievement
previously by the two groups mentioned above. Regan et al. (1996) combined 
atomic hydrogen 21~cm measurements with ionized hydrogen \ha\ Fabry--P\'erot
spectroscopy to derive a rotation curve for NGC~1530. From this they derived
a bar pattern speed of 20~\kms~kpc$^{-1}$ and used it to predict resonances 
at radii coinciding with prominent morphological features, in\-clu\-ding an outer
gaseous ring seen in \hi, an inner ring seen in \hi\ and \ha, and the nuclear
star--forming ring mentioned above. Downes et al. (1996) mapped the galaxy in 
CO and identified the major concentrations of potentially star forming
molecular gas; these are located over the central prominent inner disc
zone of radius $\sim$2~kpc
and extend out more weakly over the bar to less prominent concentrations over 
the ends of the bar. They added interferometric mapping of the central zone,
finding a circumnuclear ring structure with gas in orbits apparently
perpendicular to the elongated orbits in the outer parts of the bar. They 
relate these patterns to the predictions of resonance mo\-dels (Athanassoula
1984; Jenkins \& Binney 1994) in which gas within corotation moves along $x_1$
orbits along the bar until it reaches an inner Lindblad resonance (ILR) where it moves 
into the $x_2$ orbits perpendicular to the bar. A
neat schematic diagram in fig.~10 of Downes et al. (1996) illustrates the predicted
behaviour of the gas under these circumstances. Although the stars follow
quasi--elliptical $x_1$ orbits, in the case of gas, the shocks are produced along the bar and
specially at the ends of the bar, where orbit crowding occurs, as  was well predicted 
in Roberts et al. (1979). However,
as the gas reaches the outer limit of the zone dominated by $x_2$ 
dynamics, it suffers another shock, and is pulled through a 
right--angle, falling in again towards the nucleus. Gas on $x_1$ orbits with
an impact parameter a little larger with respect to the nucleus is less affected by the 
effects of the resonance, and proceeds along the outgoing edge of the bar,
albeit with declining velocities. It is just this kind of arrangement for 
allowing gas to fall to the centres of barred galaxies which was invoked by 
Shlosman et al. (1989) in their ``bars within bars" sce\-na\-rio as a fueling 
mechanism for the different types of circumnuclear activity: starbursts and
AGN. 

Using Fabry--P\'erot \ha\ mapping of NGC~1530, Regan et al. (1997) demonstrated
that the two--dimensional \ha\ velocity field in the bar region shows the skew
pattern predicted for $x_1$ orbits superposed on the general galactic
rotation. They went
on to combine hydrodynamical mo\-dels with their observations to compute a net inflow
rate  to the nuclear zone of $\sim 1$\Msun\ per 
year, with the re\-maining gas spraying back outwards along the major bar axis before
coupling into a new inflow cycle. Using molecular e\-mi\-ssion from CO and HCN, the former at angular 
resolution of 1.8\arcsec\ and the latter at 3.6\arcsec, Reynaud \& Downes
(1997) detected steep velocity gradients spatially confined to two\- arcs 
coupling into a ring. These arcs re\-pre\-sent shock fronts where the inflow along
the bar is being constrained to change direction and assume the $x_2$ orbital
configuration along the nucleus. These arcs correspond to two small
arm features picked out in \ha\ to the north and south of the circumnuclear
zone of enhanced star formation (see Fig.~\ref{mosaicmaps}a of the present paper). 
The points where the arcs couple to the circumnuclear 
ring are the main emitters in HCN, suggesting points of maximum mo\-le\-cu\-lar 
density. Reynaud \& Downes postulate the presence of two ILR's: one at
1.2~kpc from the nu\-cleus where the shock arcs are rooted, and one very close 
to the nu\-cleus, at the inner edge of the molecular ring. They support
these inferences using resonance theory and
their rotation curve, which is derived kinematically but supported by a 
dynamical model. In a subsequent article Reynaud \&  Downes (1999) map the same
area in two emission lines each of $^{12}$CO and $^{13}$CO, showing that the
molecular gas density is highest in the circumnuclear ring and not so high
in the arcs referred to above.

Perhaps of most interest and relevance for the study we present here
is that of Reynaud \& Downes (1998), who use their CO kinematics of the 
whole bar region in NGC~1530 to explore the relation between velocity changes
and star formation. They conclude that strong velocity
changes, which represent shocks and shear, are negatively co\-rre\-lated with 
star formation and that the distribution of star formation does not correspond fully to the 
distribution of available molecular gas, as traced by CO. Although it 
is true that the major concentration is found in the central circumnuclear 
zone where star formation is most enhanced, there is considerable CO measured
along the bar as well as at its ends, while the star formation is 
clearly weak along of the bar. Reynaud \& Downes
find ve\-lo\-ci\-ty jumps across the dust lanes from ve\-lo\-ci\-ties of order 100~\kms\ 
along the bar in one direction, to similar ve\-lo\-ci\-ties in the reverse
direction. These jumps maximize half way out along the bar from the
nucleus on both sides. They conclude that steep velocity gradients, both shear
and shock, prevent molecular clouds from condensing to form stars. One of the main aims of
the present paper is to examine this hypothesis in a little more detail using an \ha\ velocity field.

We present new high resolution velocity and intensity maps of 
NGC~1530, obtained in \ha\ emission using a Fabry--P\'erot interferometer. The instrument
provides a three--dimensional ``data cube'', akin to a conventional 
radiastronomical set of line maps, in which the emission intensity is given as 
a function of three coordinates: two positional coordinates and the
velocity. From this ``data cube''  
one goes on to compute maps in integrated line intensity, line peak velocity,
and velocity dispersion. The next step is to obtain the rotation curve of 
the galaxy and to subtract it off two--dimensionally from 
the complete observed velocity field. This allows us to obtain the
non-circular motions due to non-axisymmetric elements of the galaxy.
For NGC~1530 the dominant source of non--axisymmetry is the strong bar of length almost 28~kpc,  
which dominates the appearance of the galactic disk. We also find kinematic
signatures of the streaming motions in the sharply defined 
spiral arms. Besides we also see, in the central 5~kpc, 
evidence for a distinct kinematic structure predicted by classical resonance
theory, and already carefully observed and characterised in the studies by
Regan and coworkers, and by Reynaud \& Downes in the re\-fe\-ren\-ces given above.

As a further step we have prepared plots of the velocity gradients in the 
whole of the bar and arm structure as ima\-ged in \ha\ emission. We have related
the velocity gradient morphology to the distribution of the 
massive star formation in arms, bar, and the central zone.
Our main steps forward compared with previous studies are due to the uniformly
high angular resolution we could obtain over the full image. This has enabled us
to make a more precise de\-fi\-ni\-tion of the rotation curve, and hence 
an improved map of the non-circular velocity field, both around the main bar,
and in the central zone. From the two--dimensional display of the
non-circular velocities we have gone on to derive a map of the velocity
gradients, with a resolution better than 2\arcsec\ over the whole field of the
galaxy. Such a map has not been previously produced, so we are able to make 
progress here in characterising the geometrical and physical relationships
between velocity gradients and star formation.          
%________________________________________________________________________________________

\section{Observations and data reduction}
The Fabry--P\'erot observations of NGC~1530 were taken at the Roque de los Muchachos Observatory
with the 4.2m William Herschel Telescope (WHT) on the night of September 2nd,
1998. The TAURUS-II instrument, equipped with a TEK CCD camera was used to
observe the galaxy in the \ha\ emission line, with a plate scale of
0.292\arcsec\ per pixel,  which corresponds to a pixel scale
179~pc~arcsec$^{-1}$ at the distance of the galaxy (37 Mpc).
%The CCD was windowed to
%a total field size of 3.5\arcmin$\times$3.5\arcmin.
%720$\times$720~pixels of 0.292\arcsec\ per pixel, which correspond to a
%total field size of 3.5\arcmin$\times$3.5\arcmin.
The observations consisted of exposures of 150~s at e\-qually spaced
positions of the etalon, at intervals of 0.407 \AA, co\-rresponding
to 18.62~\kms. The to\-tal num\-ber of steps\- was 55, allowing us to scan the full 
wavelength range of the \ha\ line in the galaxy. We used the appropriately
redshifted narrow band H$\alpha$ filter as an order-sorting filter
($\lambda_{\rm c}$=6613\AA, $\Delta$$\lambda$=15\AA, corresponding to
the galaxy's systemic velocity $v_{\rm sys}$=2461~km~s$^{-1}$; de~Vaucouleurs et al. 1991; 
hereinafter RC3). 
Wavelength and phase calibration were performed using observations of
a calibration lamp (taken before and after the observations of the galaxy),
u\-sing the {\sc TAUCAL} package in the {\sc FIGARO} environment. 
The spa\-tial resolution of the resulting data set is $\sim$1\arcsec. The data set was
placed on a correctly oriented spa\-tial grid by comparing positions of point--like
features in the \ha\ Fabry--P\'erot intensity map with those in the narrow-band 
\ha\ image from Rela\~no et. al (2003b), yielding maximum uncertainties in
positions of 0.3\arcsec. The final calibrated data cube has $941\times$941
pixels $\times$ 55 `planes' in wavelength, se\-pa\-ra\-ted by 18.62 km\,s$^{-1}$. 
%=======Fig. 1 estab aqui!
We examined plane by plane the data cube and found that it 
had a considerable number of cosmic ray events. We eliminated them with the task
{\em imedit} in {\sc IRAF}, which allowed us to replace the anomalous pixels by a mean
background va\-lue. We determined which channels of the data set were free of
\ha\ line emission after smoothing the data to a resolution of
2.5\arcsec$\times$2.5\arcsec. We found 10 channels free of line emission on
the low-velocity side, and 13 channels on the high--velocity side. The
continuum was determined by fitting a linear relation to the line--free
channels, and was then subtracted from the line emission channels. 

In order to study the kinematics on different scales, the data--cube was
convolved down from the original resolution of
1\arcsec$\times$1\arcsec\ to resolutions of 
2\arcsec$\times$2\arcsec, 3\arcsec$\times$3\arcsec,  
6\arcsec$\times$6\arcsec and 10\arcsec$\times$10\arcsec. 
For each data set we obtained the total \ha\ intensity (zeroth moment),
velocity (first moment) and velocity dispersion (second moment) maps. The
procedure to obtain the moment maps is equivalent for each smoothed 
data cube so we describe the procedure followed for the full resolution 
data set only. A reader wanting a fuller descrip\-tion of the procedure can find 
it in Knapen (1998). The points with low
signal were eliminated producing a conditionally transferred data cube, in 
which values were retained only at positions where the intensity in the 
smoothed\- data cube of 3\arcsec$\times$3\arcsec\ was larger than 2.5 times 
the rms noise of the smoothed cube; pixel 
va\-lues at all other positions were set to undefined. We then removed
noise peaks outside the area where \ha\ emission is ex\-pected by setting the
pixel values there to undefined too. This was done interactively by inspecting the
high re\-so\-lu\-tion channel maps one by one and comparing them with the corresponding
and adjacent channel maps in the 3\arcsec$\times$3\arcsec\ data set. The
resulting data cube was then used to calculate the moment maps using the
{\sc moments} task in the {\sc GIPSY} package. 

{\sc moments} obtains the zeroth, first and second moments of the emission at each spatial
pixel in the galaxy and produces a corresponding map for each moment. In the
procedure, {\sc moments} takes into account the signal only when two conditions are 
met: the signal is higher than 2.5 times the rms noise of the high-resolution
cube, in at least three adjacent channels. The intensity map for the 2\arcsec\ resolution
data cube is shown in Fig.~\ref{mosaicmaps}a and the first moment (velocity) map is shown in
Fig.~\ref{mosaicmaps}b.

Although the  moments procedure is a more useful technique for studying the
global kinematics in the galaxy, it must be treated with caution 
when it is used for a detailed description of the kinematics of
individual \hii\ regions. 
The  moments procedure gives an underestimate of the
velocity dispersion of the profile (van der Kruit \& Shostak 1982), thus, if
a precision measurement of this quantity is required, a better fit to
the line profiles, using Gaussian functions, is needed.

\section{Kinematics}
\subsection{Rotation curve}

The rotation curve of the galaxy can be derived from the velocity map, which
gives the projected velocity along the line of sight in each position of the 
galaxy. Assuming that the measurements refer to a position on a single inclined 
plane, the projected velocity can be related to the rotation velocity in 
the plane of the galaxy by this expression:
\begin{equation} 
v(x,y)=v_{\rm sys}+v_{\rm rot}(r){\rm cos}(\theta){\rm sin}(i)
\label{eqvrot}
\end{equation}
where $v(x,y)$ denotes the radial velocity at the sky coordinates $x$ and $y$,
$v_{\rm sys}$ the systemic velocity, $v_{\rm rot}$ the rotational velocity,
$i$ the inclination angle of the normal to the galaxy
plane to the line of sight, and $\theta$ the azimuthal distance from the
major axis in the plane of the galaxy, related to the rotation centre ($x_{0}$, $y_{0}$)
and the position angle of the galaxy, PA. 
We derived the rotation curve following the procedure described by Begeman~(1989), 
where the rotation curve is determined 
by fitting tilted rings to the observed velocity field. 
For each ring independently, the parameters  
$v_{\rm sys}$, $v_{\rm rot}$, $i$ and $x_{0}$, $y_{0}$ and PA are obtained making a 
least--squares fit to Eq.~\ref{eqvrot} using the observed velocity field $v(x,y)$. 

We discarded the points within a certain free--angle $\theta_{\rm max}$ around 
the minor axis since they carry less information about the underlying circular 
velocity of the galaxy and the values obtained in this zone are less accurate due 
to projection effects. In the case of NGC~1530 a value of $\theta_{\rm max}=45\deg$ 
was used to avoid the non-circular motions along the bar. Besides, we give more 
weight to positions close to the major axis using a
weighting function proportional to $|{\rm cos}~\theta|$ in the fit. The procedure used 
to obtain each of the parameters which characterise the rotation curve is the following.  

The parameters $v_{\rm sys}$, $x_{0}$ and $y_{0}$ depend on the symme\-try of
the velocity field and are the same for each ring.
We obtained the position of the kinematic centre
$(x_{0},y_{0})$ using the velocity map of 6\arcsec\ resolution. We took
annuli of 3\arcsec\ width, fixed values for the heliocentric systemic ve\-lo\-city at 2461\kms\
(RC3), inclination angle at $i=50\deg$ (a mean between the values of
Downes et al. 1996 and Regan et al. 1996) and position angle  at
188\deg\ (the value found by Regan et al. 1996) and obtained $(x_{0},y_{0})$ using 
Eq.~\ref{eqvrot}. We found a large scatter in the values of the
kinematic centre for annuli far from the centre of the galaxy. This is due 
to the paucity of \ha\ 
emission in the disc of the galaxy outside the arms and to the fact that
lower re\-so\-lution data cubes are more spatially affected by the smoothed
non-circular motions along the bar in the data cubes. The mean value of
$(x_{0},y_{0})$  for the rings inside 12\arcsec\  corresponds to the coordinates 
$\hmsd 4h23m26.94s$($\pm$0.02$^{\rm s}$) $\dms 75d17m44.0s$($\pm$0.1\arcsec), J2000. We checked this value
using the same procedure for the velocity maps of lower and higher (10\arcsec\ and 
2\arcsec) resolution, and found that the differences in the rotation centre from 
the value adopted were inside the error bars. 

\begin{figure}
\begin{center}
\resizebox{9cm}{!}{\includegraphics{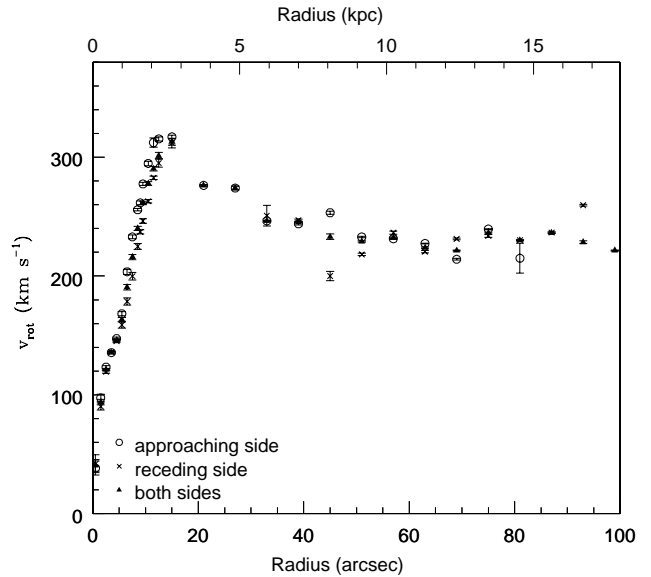}}
\parbox[e]{8.5cm}{\caption{\ha\ rotation curve of the disc of NGC~1530,
for the receding and the approaching sides of the galaxy separately. The
agreement between the two sides shows the 
symmetry of the velocity field and the reliability of the rotation centre.}
\label{curva2}}
\end{center}
\end{figure}

The systemic velocity was obtained from the 10\arcsec\ re\-so\-lu\-tion velocity
map with rings of width 5\arcsec. The centre position was fixed to the value found previously, 
and the inclination and position angle were fixed to 
the same va\-lues used to derive the rotation centre. The mean value for  
$v_{\rm sys}$ in all the rings is $2466\pm4~\kms$, where the error is
the standard deviation from the mean. This va\-lue
is in agree\-ment with the value found by Reynaud et al. (1997) of 
$v_{\rm sys}=2466\kms$ for CO data, but di\-ffers from that  
obtained by Regan et al. 1996 in \ha, $v_{\rm sys}
%\footnote{\rm The velocities in this paper are 
%heliocentric velocities. To obtain local standard of rest velocities add
%5.7~\kms.}
= 2447\pm2~\kms$.
A way to test whether the rotation centre and the systemic velocity are well 
determined is shown in Fig.~\ref{curva2}, 
where we have plotted rotation curves for the receding and a\-pproaching halves 
of the galaxy derived separately u\-sing the rotation centre and systemic velocity 
found above. The coincidence of the velocity points for both sides, es\-pe\-cially in the 
central zone of the galaxy, provides evidence of 
the symmetry of the velocity field and the validity of the va\-lues for the rotation centre and 
systemic velocity. We also derived the rotation curves for the re\-ce\-ding and a\-pproaching
halves with the rotation centre found in Regan et al. (1996), but in that
case we found discrepancies which appeared unphysically large. It is the better
resolution of our data (1\arcsec, while Regan et al. (1996) has a net
resolution of 4\arcsec) which  allows us to obtain a more accurate position 
for the rotation centre. 

\begin{figure}
\begin{center}
\resizebox{9cm}{!}{\includegraphics{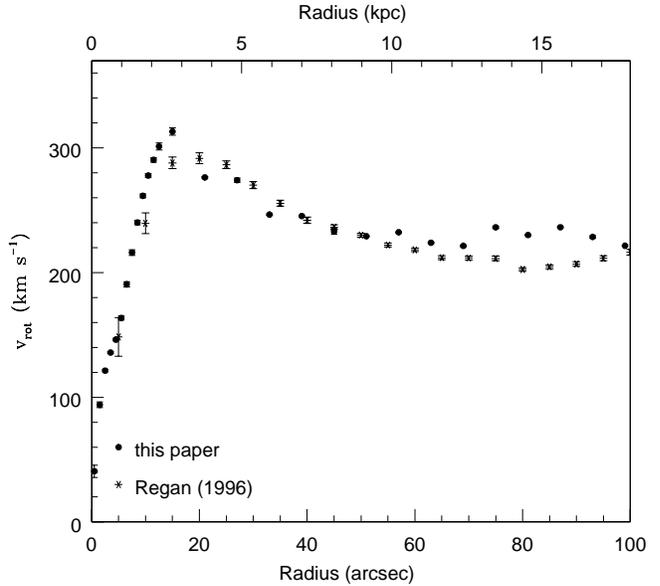}}
\parbox[e]{8.5cm}{\caption{\ha\ rotation curve of the disc of NGC~1530, 
up to ~18 kpc from the galactic centre. The rotation curve was obtained by fitting tilted rings to the
observed velocity field in Fig.~\ref{mosaicmaps}b. Also shown is the rotation curve obtained by Regan et al. (1996).}
\label{curva}}
\end{center}
\end{figure}

Finally, we obtained the rotation curve using the velocity map at 2\arcsec\
resolution in Fig~\ref{mosaicmaps}b. For the internal part of the galaxy, out to 
12.5\arcsec, we used rings of width 1\arcsec, and rings of width 6\arcsec\ for 
larger radii. The rotation curve obtained in this way is shown 
in Fig.~\ref{curva}, where the PA and the inclination were fixed at 
188\deg\ and 50\deg\ respectively. Fig.~\ref{curva} also displays the 
rotation curve from Regan et al. 1996, which is similar 
to that found here, but we obtain higher values of the rotation velocity in 
the inner part, 10\arcsec--15\arcsec,
and also in the 70\arcsec--90\arcsec radial range.
In order to check the PA adopted, we left it as a free parameter in the fit, 
finding that the values given by the program for most of the annuli are close 
to 188\deg\ and that the rotation curve did not change significantly from the curve with a fixed PA. The 
adopted inclination angle $i=50\deg$ was also checked. We fitted $v_{\rm
  rot}$ and $i$ with Eq.~\ref{eqvrot}, keeping the other\- kinematic parameters fixed, and 
found a mean value of $i=(53\pm8)\deg$, in good agreement with the value adopted.

\begin{figure}
\begin{center}
\resizebox{9cm}{!}{\includegraphics{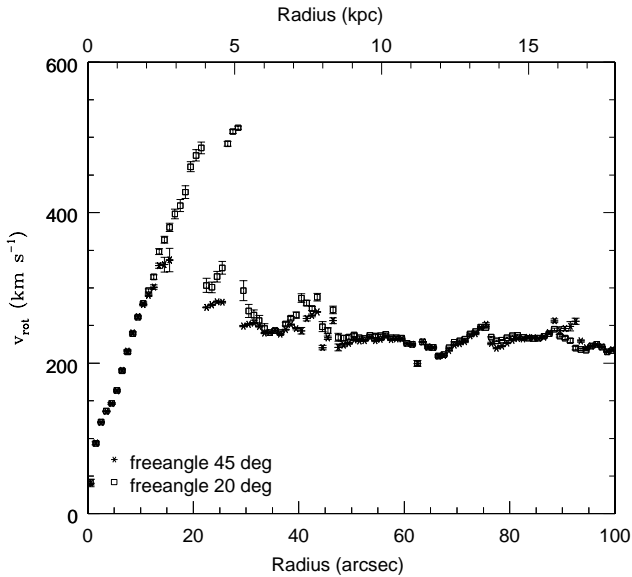}}
\parbox[e]{8.5cm}{\caption{\ha\ rotation curves of the disc of
    NGC~1530 obtained with different 
 ``free--angles'' and with rings of 1\arcsec\ width. For free--angles less
 than $\sim$ 30\deg\ there is an obvious contribution from the
 non--rotational motion along the bar in the mean calculated rotational
 velocity that can ``contaminate'' the rotation curve of the galaxy.}
\label{curvafreeangle}}
\end{center}
\end{figure}

To test the importance of a careful definition of the the free--angle we 
reduced its value to 20\deg\ and rederived the rotation curve. As seen in
Fig.~\ref{curvafreeangle} this produced ``rotational velocities'' rising to
values of over 500~km\,s\me, as there is
an obvious contribution from the non--rotational motion along the bar projected
into the line of sight. As we will see below the amplitude of the gas motion 
along the bar is over 100~km\,s\me, and it is easy to see that this will
yield the unrealistically high velocities shown in
Fig.~\ref{curvafreeangle}. Fortunately, the bar orientation in NGC~1530 does 
not make it too difficult to isolate and avoid the emission from gas which contains 
a significant non--circular velocity component, but Fig.~\ref{curvafreeangle} offers a warning 
to those using maps of much lower resolution in similar circumstances.

\begin{figure}
\begin{center}
\resizebox{9cm}{!}{\includegraphics{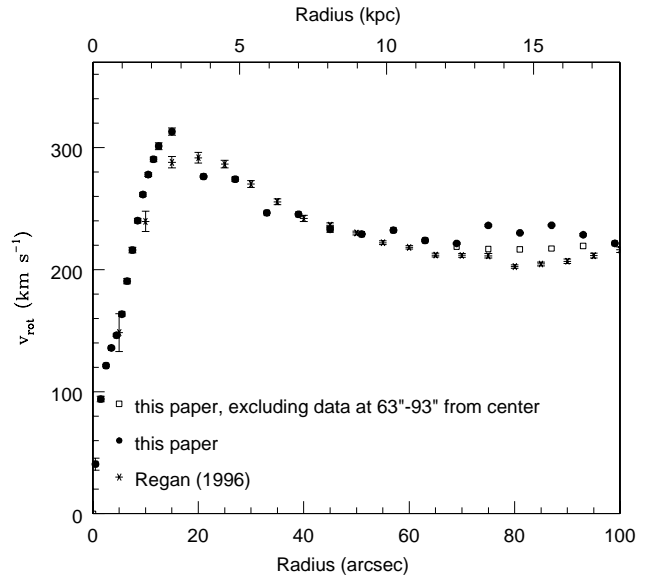}}
\parbox[e]{8.5cm}{\caption{\ha\ rotation curve of the disc of
    NGC~1530. This rotation curve has been obtained by  
subtracting the effect of the streaming motions in the spiral arms of the galaxy 
from the mean rotation curve shown in Fig.~\ref{curva} (see text for further details).}
\label{crmodified}}
\end{center}
\end{figure}

\subsection{The residual velocity map}
After deriving the mean rotation curve, an easy way to explore non--circular 
motions within the galactic disc is to make a two--dimensional projection of
this curve and subtract it from the observed velocity field. 
 The two--dimensional projection of the initial rotation curve is seen in 
Fig.~\ref{mosaicmaps}c, where we have superposed an isophote map of intensities to show
the positions of the emitting components. The projection has classical symmetry,
but shows a few small features: ``wiggles" which occur at some three quarters
of the radial distance from the centre to the edge of the field. These can be
explained as effects of streaming motions in the spiral arms, which have not
been subtracted adequately from the mean rotation curve.  

While it is difficult 
to get an exact separation of rotation and streaming motions, we have made a 
first a\-pproxi\-mation by assuming that the true rotation cur\-ve is\-~smooth.  
Applying a second order polynomial fit to the part that is observed as smooth, 
i.e. between 27\arcsec\ and 63\arcsec\ from the centre, we then extrapolate this to cover 
the range between 63\arcsec\ and the edge of the field at 99\arcsec\ from the centre, 
and go on to combine it with the original curve out to 63\arcsec\ radius, to give the final
``true" rotation curve. The result of this procedure is shown in Fig.~\ref{crmodified}, 
where we have compared our original rotation curve with the new curve and
also with the curve by Regan et al. (1996). Our procedure has isolated the 
wiggles between 63\arcsec\ and 93\arcsec. 

We can then proceed to produce a new
two--dimensional projected model for the rotation in which the wiggles are no
longer present,  and to subtract off this new rotation 
model from the original velocity field, yielding a resulting map of residuals 
as shown in Fig.~\ref{mosaicmaps}d.  There are clearly ordered fields of non--circular velocity 
both around the bar, and in the arms, but we can check on the mo\-del subtraction by 
examining the residuals of the regions which are in the disc outside the arms and bar. 
These in general show, as they should if the rotation curve subtraction is valid, net residual 
velocities close to zero: less than $\sim$10~km~s\me, which is
close to the precision limit of the observations. We can now see clearly the 
streaming motions across the arms in Fig.~\ref{mosaicmaps}d. These have
projected values between 10 and 30~km~s\me, with similar amplitudes in both arms.

\subsubsection{The residual velocities along the bar}

Of most striking in\-te\-rest is the fairly complex non--circular
velocity field associated with the bar. Approaching velocities of up to 
%just over 
120~km~s\me\ are observed along the north side of the bar, and re\-ce\-ding 
velocities of the same amplitude along the south side. These values, and the
implied vec\-torial direction in elongated quasi--elliptical orbits around the 
bar with long axis parallel to the bar axis, are in line with classical predictions 
of the dynamical effects of a stellar bar on the rotating gas 
which date back to Huntley et al. (1978) and Roberts et al. (1979) as well as
to more recent studies by Athanassoula (1992). 
Most of the observational work has been performed either in 
\hi\ or more recently in CO emission. There have been relatively few \ha\ 
studies of this phenomenon, for two probable reasons: slit spectroscopy is 
costly in time if more than a couple of position angles are taken (one good
early study in this mode using six slit positions on NGC~1512 was 
published by Lindblad \& Jorsater 1981); and \ha\ emission is much patchier 
than either \hi\ or CO.  However, as we can see from the data in the present 
article, the use of a scanning Fabry--P\'erot for 
this work permits full two--dimensional coverage at an angular resolution
better than that of essen\-tially all \hi\ or CO measurements, which can compensate 
to an interesting degree for the intrinsic patchiness of the observed field.  

In Fig.~\ref{mosaicmaps}d there are several complexities in the non--circular velocity field 
which are worth pointing out.  Taking the north side of the bar we can see a zone with high
bar--streaming velocities, of order~120 km~s\me\ in a direction approaching us 
to the east of the centre, and another region, more restricted in size and velocity 
range, with si\-mi\-lar velocities to the west. To the south of the bar we can see a 
diametrically symmetric flow, with high re\-ce\-ding velocities, again of order 120~km~s\me\ 
receding on west side, and two smaller zones with similar velocities to the east. 
It is tempting to consider these flows as direct evidence for motion aligned with 
the bar and terminating in zones of shock towards the ends of the bar. We 
will see later that there is considerable evidence in favour of this, and of 
the direct causal links between properties of these flows, and both the
stimulation and inhibition of massive star formation.

Examining Fig.~\ref{mosaicmaps}d, the question arises as to how the flow can apparently be
stopped half way along the bar, yet continue on beyond the central zones as 
described above. This appearance does not give a true picture of the velocity 
field and is essentially due to the patchiness of the emission. However
taking the latter into account we can interpret the observations in terms
of a resonance model of the type illus\-tra\-ted in Downes et al. (1996) 
to interpret their measurements in CO, based on models incorporating an ILR
 to a main galactic bar by Athanassoula (1992).  
Qualitatively, these models show how the bar dynamics produces orbits termed 
$x_1$, whose major axes are aligned with the major axis of the bar, but that the presence of an inner
Lindblad resonance within  the bar structure (not far from the nucleus) 
implies that gas will move on the perpendicular $x_2$ elliptical orbits 
within the ILR. The behaviour of a particular gas cloud in an $x_1$ orbit will then depend 
critically on its impact parameter with respect to the nucleus, i.e. on the 
length of its orbital semi--minor axis. Gas on orbits close to the bar major 
axis will be dragged into $x_2$ orbits as it passes close to the nucleus, and
will not escape along the bar. However, gas 
on orbits well away from the major axis will remain essentially on those orbits
until it reaches the far end of the bar, where it will be braked and shocked. 
Gas on intermediate orbits will swing in towards the bar major axis, where
in turn it will also be braked and shocked. In Fig.~\ref{orbits} we have 
illus\-tra\-ted these  types of orbital behaviour, with a representative set
of five trajectories on either side of the bar. Gas mo\-ving on the innermost 
trajectory will be pulled into an $x_2$ orbit, while gas on the others will end
up impinging on the bar progressively further along, as their
initial impact parameters increase. Because the velocity vector is observed 
only where star formation is occurring, we cannot follow these trajectories 
along their lengths, but see clearly their initial and final points in the 
residual velocity field.   

\subsubsection{Residual velocity cross--sections along and perpendicular to the bar}
\begin{figure*}
\resizebox{0.9\textwidth}{!}{\vspace{-3.5cm}\includegraphics{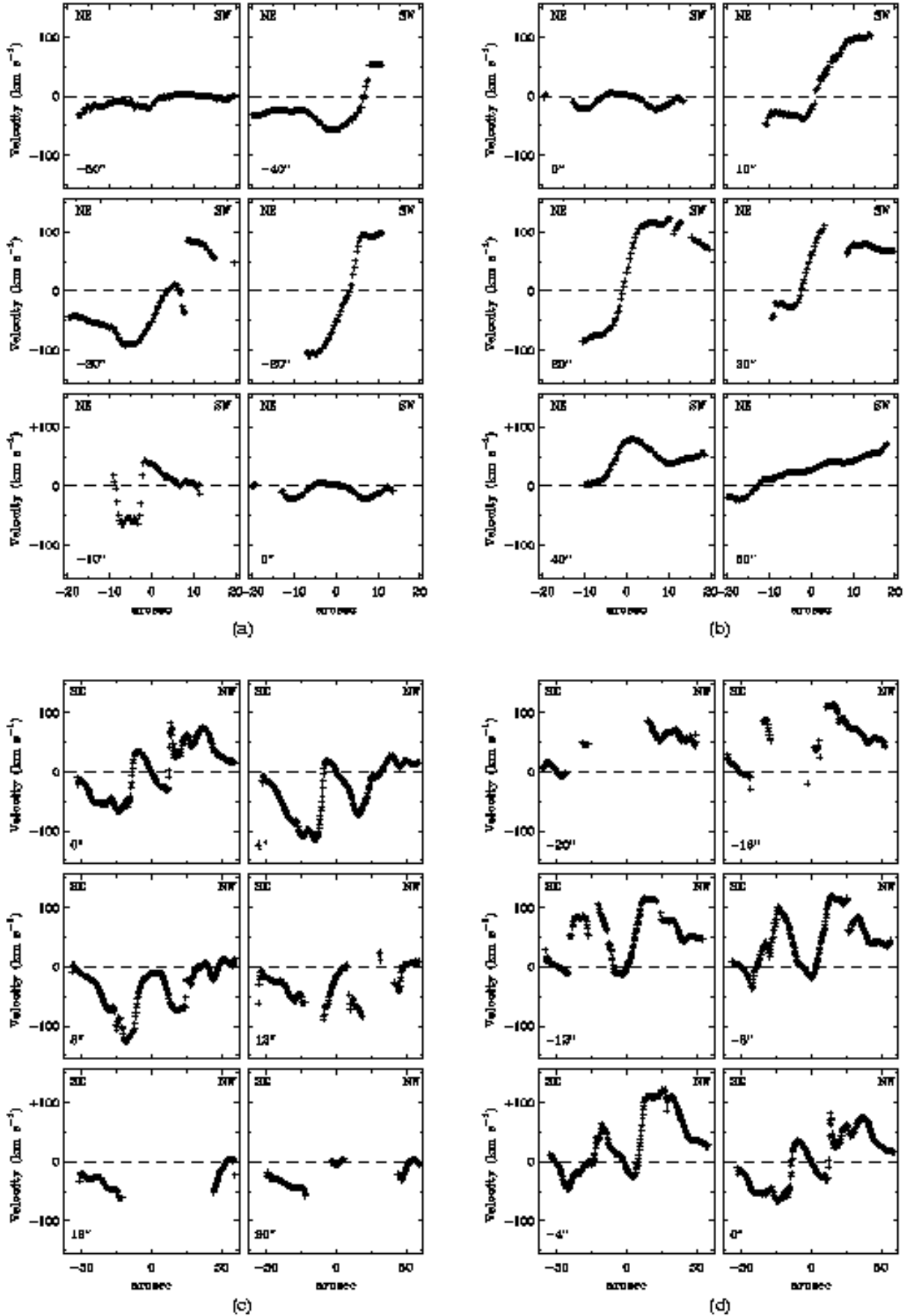}}
\parbox[e]{\textwidth}{\caption{
Profiles of the residual velocity perpendicular to the bar of the galaxy: 
to the SE (a) and to the NW (b),
and parallel to the bar: to the NE (c) and SW (d). The
profile tracks are located towards the 
left, right, top and bottom  of Fig.~\ref{perfilresiduos} in panels (a), (b), (c) 
and (d), respectively. 
The distances of the  profile tracks to the galaxy kinematic centre  are 
indicated in the bottom--left hand corner of each plot. } 
\label{perfiles}}
\end{figure*}

In Figs~\ref{perfiles}a, \ref{perfiles}b, \ref{perfiles}c and~\ref{perfiles}d 
we show profiles in residual velocity perpendicular and parallel 
to the bar direction (PA=116\deg). The positions and directions of the
cross-sections chosen are plotted over the residual velocity map in 
Fig~\ref{perfilresiduos}. For profiles perpendicular to the bar and lo\-ca\-ted 
where the bar connects 
with the spiral arms (-40\arcsec\ in Fig.~\ref{perfiles}a and 40\arcsec\ in
Fig.~\ref{perfiles}b) the absolute value of the re\-si\-dual velocity along the
line of sight is $\sim$70--80\kms. The velocities rise as we approach the
centre of the bar and at $\pm$30\arcsec\ 
are $\sim$100\kms. We note particularly the symmetry of the profiles on
either side, at $\pm$20\arcsec\ of the profile through the nucleus. We can
also see very clearly the steep velocity gradients at these positions. 
In the profiles parallel to the bar (Fig~\ref{perfiles}c and
Fig~\ref{perfiles}d) there is also evidence of rapid velocity changes. In
particular the profiles at 4\arcsec\ and 8\arcsec\ in Fig~\ref{perfiles}c, and 
-12\arcsec\ and -8\arcsec\ in Fig~\ref{perfiles}d, show very sharp velocity
gradients. These gradients mark a combination of braking and projection effects:
braking effects  as the gas flow along 
the bar meets the central gas concentration, and projection effects as the
flow vector switches direction from nearly along the line of sight (along the 
bar) to nearly perpendicular to the line of sight (as the gas is pulled into circumnuclear orbits).
A result of examining these profiles was to induce us to make a more
detailed map of the velocity gradients, which we will show and discuss below.

\subsection{Residual velocity gradient maps}
\label{sectiongrad}

Following up the indication of the presence of steep velocity gradients
in Figs.~\ref{perfiles}a, \ref{perfiles}b, \ref{perfiles}c and~\ref{perfiles}d,
we opted to produce a full map of the velocity gradients
in the system, taking advantage of our excellent angular resolution. To make 
sure that the steep gradients were not arte\-facts of our data processing we 
first used the 2\arcsec\ resolution data cube, and made position--velocity  diagrams at those
points where Figs.~\ref{perfiles}a, \ref{perfiles}b, \ref{perfiles}c, \ref{perfiles}d 
indicated that we ought to find rapid velocity 
changes. This exercise was non-trivial, sin\-ce the steepest gradients occur in
zones of low \ha\ intensity, for reasons which we will discuss further
below. The dia\-grams are shown in Fig.~\ref{posvelgrad}, where we can see at once that the 
velocity gradients are real and very steep, at the positions chosen.
We go on to derive maps in velocity gradient, selecting 
two directions, parallel and perpendicular to the bar, to optimize the 
information about the gas flows induced by the gravitational perturbation 
of the bar. The gradient for a given direction was derived in a straightforward
way as follows: the velocity residual map was displaced by one pixel in 
that direction, and sub\-tracted off from the original map; this procedure was
repeated in the opposite direction, and the gradient was found by co--adding the 
absolute values for the two maps, and normalizing the result, dividing by two.
The same procedure was followed for both of the selected directions, yielding 
two independent velocity gradient maps. The whole map of residual velocity gradient 
in the direction perpendicular to the bar is shown in Fig.~\ref{grad+V}a, while in 
Figs.~\ref{mosaico_barra}d and \ref{mosaico_barra}c we show maps of the gradients
in residual velocity, measured in the directions perpendicular to the bar and
along the bar respectively, in the zone of the bar.
The regions of steep gradient show up very
clearly, and reach high values ($\sim0.35\kms$pc\me). Near the centre of the bar we can pick out 
elongated regions where the motions along the bar axis are in counterflow, 
so that an amplitude of over 100~\kms\ in one direction rapidly gives way 
to the same amplitude in the opposite direction. The measured gradients imply 
that this occurs over distances of a few hundreds of parsecs, close to the limit of 
direct detection, so that there may in fact be much steeper shocks than this.
Our decomposition of the gradients into these two perpendicular directions
does not effectuate a very clean separation between different intrinsic
dynamical e\-ffects, above all because the observations give the projected
components along our line of sight, and projection effects are not removed.
This would be difficult to do correctly without a previous underlying model, 
which would assume too much about the directions of motion to be reliable, so 
we prefer to leave the observations as they are. In general, zones of high
velocity gradient in one direction also show 
\onecolumn
\newcounter{figmas}
\addtocounter{figmas}{7}
%%%%%%%%
\renewcommand{\thefigure}{\arabic{figmas}}
%%%%%%%%
\vspace{-3cm}
\begin{figure*}
\begin{minipage}[b]{0.5\textwidth}
\vspace{0.5cm}
\centering
\includegraphics[width=9.5cm]{ms4084.f7.lr}
\caption{Lines tracing the cross--sections perpendicular and
    pa\-ra\-llel to the bar of NGC~1530 represented in Figs.~\ref{perfiles}a
    to~\ref{perfiles}d over a portion of the residual  
velocity map.}
\label{perfilresiduos}
\end{minipage}
\addtocounter{figmas}{1}
\hspace{0.3cm}
\begin{minipage}[b]{0.5\textwidth}
\centering
\vspace{-1.0cm}
\includegraphics[width=9.0cm]{ms4084.f8.lr}
\caption{Schematic representation of the gas orbits along the bar on the
    residual velocity map, illustrating the dependence of the gas
    trajectories on orbit semi-minor axis.}
\label{orbits}
\end{minipage}
\addtocounter{figmas}{1}
\begin{center}
\begin{minipage}[b]{\textwidth}
\centering
\includegraphics[width=\textwidth]{ms4084.f9.lr}
\caption{{\bf a)} Velocity gradient map perpendicular to the bar. Zones of maximum
  shear are clearly seen (cf. the dust lanes in Fig.~\ref{grad+V}b).
{\bf b)} Image of NGC~1530 in the optical {\it V} band obtained with 
the KPNO 4--meter Mayall telescope (NOAO/AURA/NSF).}
\label{grad+V}
\end{minipage}
\end{center}
\addtocounter{figmas}{1}
\end{figure*}
\twocolumn  %*******Poner \onecolumn en version referee
\begin{figure}
\begin{center}
\resizebox{8cm}{!}{\includegraphics{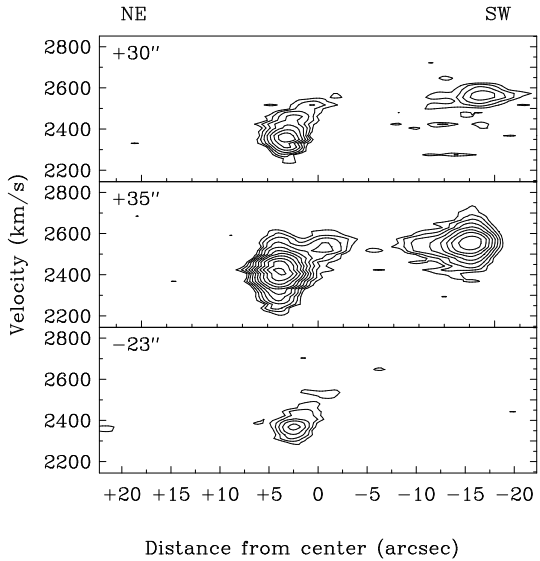}}
\parbox[e]{\columnwidth}{\caption{Position--velocity diagrams taken along
    cross--sections perpendicular to the centre line of the bar. Distances
    from the nucleus are shown in the top left hand corner. Distances are positive 
    towards the NW and negative towards the SE.}
\label{posvelgrad}}
\end{center}
\addtocounter{figmas}{1}
\end{figure}

\begin{figure*}
\begin{center}
\resizebox{\textwidth}{!}{\includegraphics{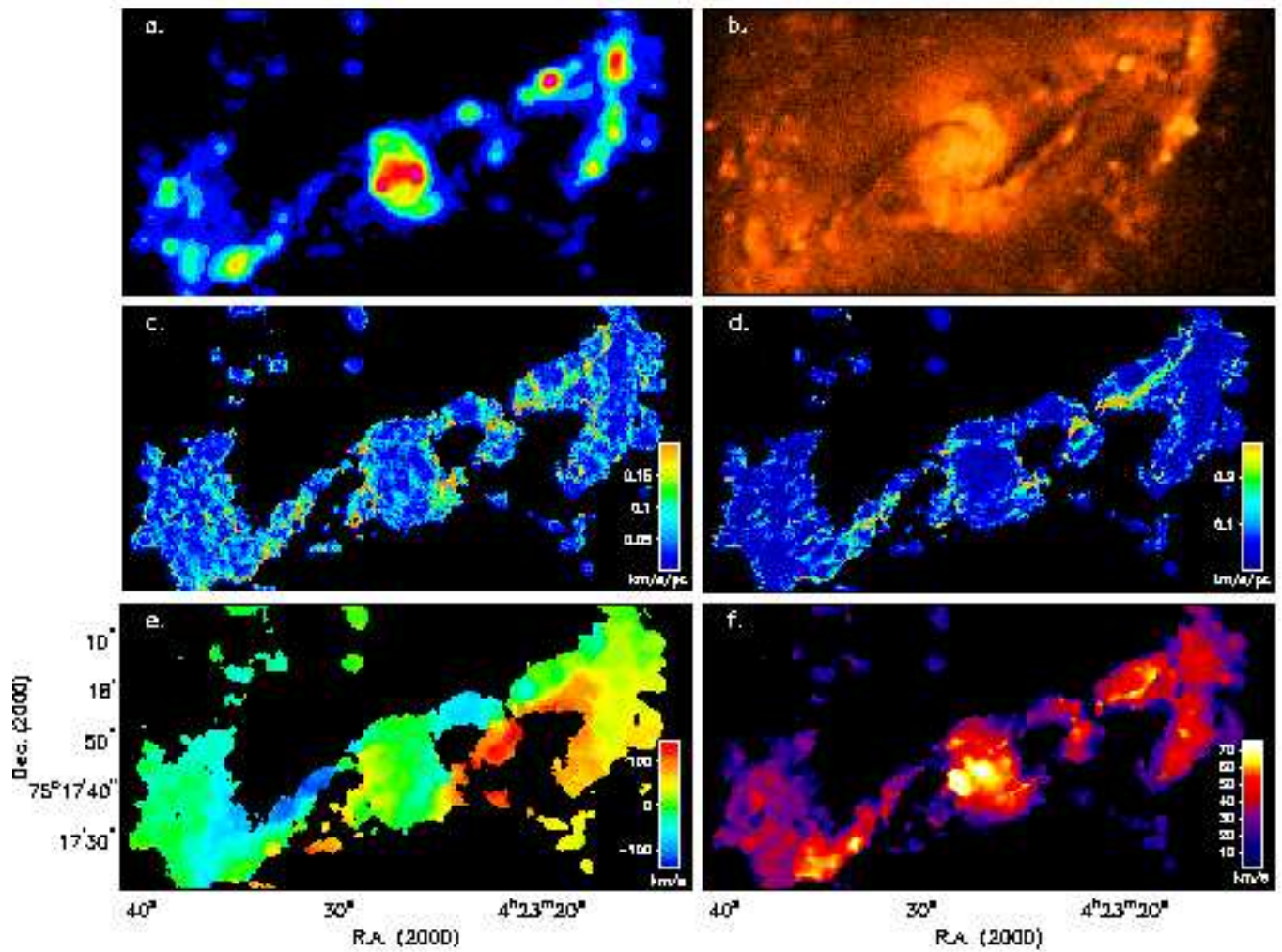}}
\parbox[e]{\textwidth}{\caption{{\bf a)} Intensity map of the \ha\ emission in 
zone of the bar.  {\bf b)} {\it V} band image. {\bf c)} Velocity gradients parallel
to the bar. {\bf d)} Velocity gradients perpendicular to the bar. {\bf e)} 
Non-circular residual velocity map. {\bf f)} Non-thermal velocity dispersion map. 
All images (a to f) show the same zone of the galaxy.}
\label{mosaico_barra}}
\end{center}
\addtocounter{figmas}{1}
\end{figure*}
\noindent In 
high gradients in the
perpendicular direction, though the details differ on very small scales. Fig.~\ref{grad+V}b we have
also included an image of NGC 1530 in the {\it V} band. The morphological
and positional agreement between the dust lanes and the zones of maximum 
velocity gradient perpendicular to the bar illustrated in this figure is
striking. This result may not be qualitatively unex\-pec\-ted, since various
models of gas flows and shocks in barred galaxies do predict that gas and
dust densities should be elevated in the shocked zones. However, to our knowledge,
this is the first presentation of a di\-rect one--to--one ma\-pping of the velocity 
gradients in shocks and the dust lanes along a bar with comparable angular resolution.

Comparing these velocity gradient maps, and in par\-ti\-cular the map
of gradients perpendicular to the bar, Fig.~\ref{grad+V}a or \ref{mosaico_barra}d,
with the total intensity map in Fig.~\ref{mosaicmaps}a
we can see clearly that the star formation is peaking in zones where the
velocity gradients are low. There is a very good anticorrelation here, which 
holds over the whole of the face of the galaxy: in the arms, along the bar, and
in the central zone. In general, when comparing the non--circular velocities in 
Fig.~\ref{mosaico_barra}e with the velocity gradients in Fig.~\ref{mosaico_barra}c,d, 
we find high velocity gradients associated spatially with significant non--circular velocities,
but there are two specific zones where this is clearly not the case. Along
the NW side of the bar there are two bright star--forming regions, the less
intense of which is situated within gas with a pro\-jected residual flow vector
of amplitude $\sim$100~\kms, as seen in Fig.~\ref{mosaico_barra}e. This same zone, 
is clearly in a velocity gradient minimum
as seen in Figs.~\ref{mosaico_barra}d and~\ref{mosaico_barra}e. A second example is the
intense star for\-ming zone where the bar meets the NW arm, where the velocity
gradients are low, but the net velocity vector has an amplitude of
$\sim$30~\kms. Both of these are regions with strong star formation, and low velocity 
gradient but relatively high re\-si\-dual velocity. We will discuss the 
implications of these correlations and anticorrelations in Sect.~4.   

\subsection{The velocity dispersion} 
The second moment of the intensity--wavelength distribution represents the
velocity dispersion of the emission line observed. Since we are interested in 
the dispersion due to the turbulent (non-thermal) motions of gas, we
must extract it from the observed \ha\ emission line profiles 
taking into account  effects which also
contribute to broaden the observed line shape. 
These are the natural \ha\ emission line
broadening, thermal broadening and ins\-tru\-mental broadening. 
We subtracted these components in quadrature
from each point of the observed velocity dispersion map, to 
obtain a new map of the dispersion due to the non--thermal 
motions of the ionized gas. For further details on the 
procedure see Rozas et al. (1998, 2000).
In Fig.~\ref{mosaico_barra}f 
we show the map of non--thermal velocity dispersion in the zone of the bar. 
In general there is a close correlation between high 
surface brightness (as seen in Fig.~\ref{mosaico_barra}a) and high 
velocity dispersion, as expected knowing that much
of the measured turbulence is stimulated by the energy output of young massive
stars, which also give rise to the \ha. There are some zones, notably 
towards the ends of the bar, where this agreement is less clear, 
i.e. where the  ratio of the \ha\ surface brightness to the measured 
dispersion is much lower than for other star forming zones. We attribute 
this to the fact that our measurements of dispersion cannot avoid including
 components of steep  localized velocity gra\-dients which are particularly 
important where the arms break away from the bar. As we will see in 
Sect.~\ref{barCS}, there is a clear anticorrelation
between the local star formation rate and the gas velocity gra\-dient, and the
observations of dispersion alone would not easily pick this up. 

\section{Relationships between the velocity field and the local star formation rate}

\subsection{Bar cross--sections in velocity and \ha\ surface brightness}
\begin{figure*}
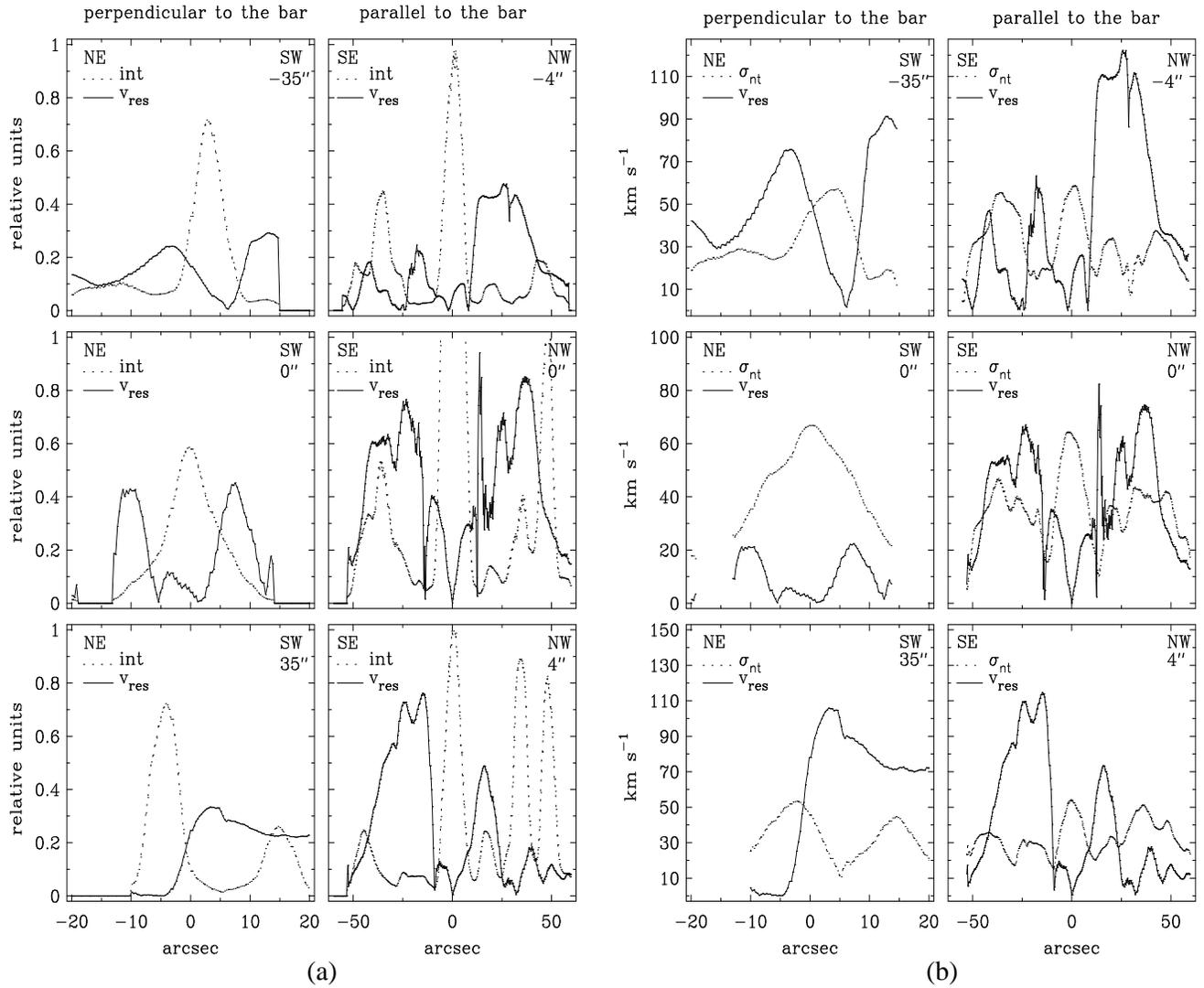

\begin{center}
\resizebox{\textwidth}{!}{\includegraphics{ms4084.f12a}\includegraphics{ms4084.f12b}}
\parbox[e]{\textwidth}{\caption{{\bf a)} Profiles of the normalized surface
brightness in \ha\ emission and residual velocity parallel and perpendicular 
to the bar. The distance
of each profile track from the kinematic centre is indicated in the top right
corner of each plot. Positive distances in the perpendicular and parallel 
profiles are respectively towards the NW and NE in Fig.~\ref{perfilresiduos}. {\bf b)} Profiles of 
the absolute residual    velocity and nonthermal velocity dispersion parallel and perpendicular to
the bar. The tracks of the profiles are the same as those in Fig.~\ref{perfilesA}a.}
\label{perfilesA}}
\end{center}
\addtocounter{figmas}{1}
\end{figure*}

To examine more closely the relation between the cu\-rrent star formation
rate as measured in \ha\ surface brightness, and the velocity field of the 
gas in the region of the bar, we obtained a series of cross--sections in 
which the re\-si\-dual velocity can be compared directly with 
the \ha\ surface brightness, as displayed in Fig.~\ref{perfilesA}a, and 
in Fig.~\ref{perfilesA}b with the velocity dispersion. 
In Fig.~\ref{perfilesA}a we show a set of these cross--sections
perpendicular to and parallel to the bar, comparing surface brightness with
residual ve\-lo\-cities, i.e. non--circular velocity. As a general feature of the
cross--sections it is clear that there is an anti-correlation between \ha\
intensity and residual velocity\footnote{To make for a
homogeneous comparison we have plotted as positive the amplitude of the
velocity vector without reference to its direction, while from the map in
Fig.~\ref{mosaicmaps}d it should be clear  that the
velocities are in opposite directions along opposite sides of the bar.}. We
can make a plausible hypothesis that suggests two distinct phe\-no\-me\-na, one of 
which inhibits star formation and the other of which helps to produce it,
which combine to yield this general result. Around the bar, the long
elliptical orbits should have (see e.g. Huntley et al. 1978) velocity
vectors around the ellipse which are high close to the bar major axis, and
fall further and further from the axis. In this 
configuration the maximum residual velocity around the bar should be 
close to the zone of maximum velocity shear. The conditions which determine the stability 
of clouds for global star formation in galaxy disks have been shown by
Kennicutt (1989), following Toomre (1964), to depend on the relationship
between the gas surface density and the velocity dispersion in the plane.
However,
this general condition should be affected by the presence of velocity shear, 
which would act against the condensation of massive gas clouds to form 
stars, as suggested by Reynaud \& Downes (1998). Therefore, in the global motion
around the bar we might expect to find a lower star formation rate in zones
of high velocity. However, within this global framework, certain shocked zones
are pre\-dicted to be places where star formation is accelerated (see e.g. 
Elmegreen et al. 2002), and where the global motion is locally strongly
braked. This will give rise to zones of low velocity and high star formation
rate which also contribute to a negative correlation between gas velocity and
star formation rate, but with variations over more compressed spatial
scales. The global anti--correlation is well seen in all the cross-sections
of Fig.~\ref{perfilesA}a, where peaks in the \ha\
surface brightness nearly all coincide with troughs in the residual
velocity. Local effects are especially well seen in the cross--section through the nucleus
parallel to the bar, which cuts two shock fronts; near these the star formation 
rate is low, but in the radial galactocentric range between them it is very high. These
relationships are also seen in Fig.~\ref{perfilesA}b, where we have made
cross--sections along the same tracks, perpendicular to and parallel 
to the bar, but where we have plotted the velocity dispersion instead of the
\ha\ surface brightness. Qualitatively the graphs are very similar, with a
similar set of anticorrelation patterns. The main difference is that the
velocity dispersion peaks tend to be broader than those in 
surface brightness, reflecting the turbulent velocity structure within
individual ultraluminous \hii\ regions (there is a clear tendency for 
the brighter regions to have the biggest turbulent widths; see e.g. Rozas et. 
al 1998; Rela\~no et al. 2003b, in preparation). We should also note  
velocity gradients on di\-ffe\-rent scales, in particular gradients in the global
velocity pattern can contribute to an observed local value of the velocity dispersion, 
while there is no
analogous effect for surface brightness. There is quantitative material in
these data which can contribute to theoretical modelling of sce\-na\-rios for massive 
star formation, though this is beyond the direct scope of the present article.

\subsection{Bar cross--sections in velocity gradient and \ha\ surface 
brightness}
\label{barCS}

\begin{figure*}
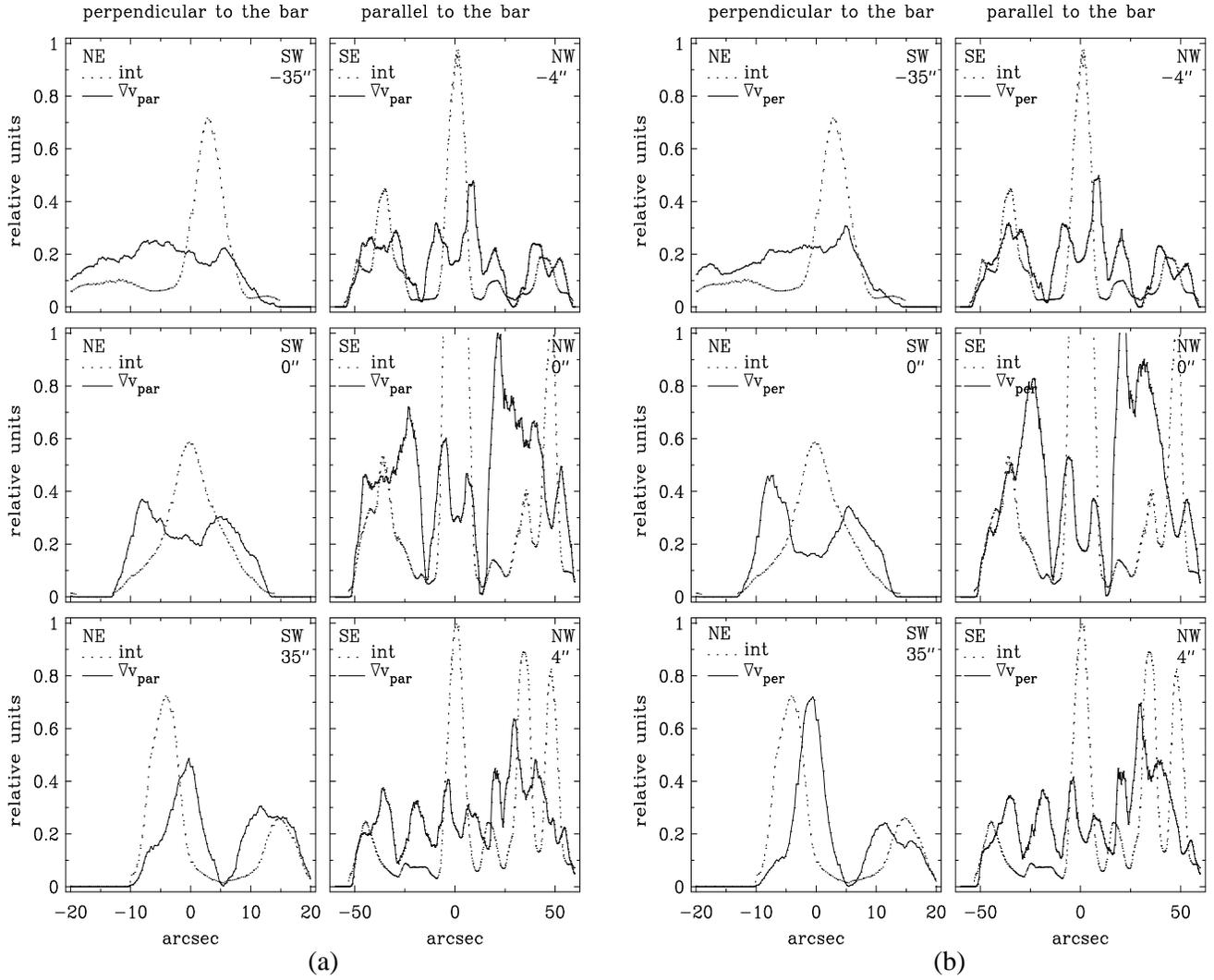

\begin{center}
\resizebox{\textwidth}{!}{\includegraphics{ms4084.f13a}\includegraphics{ms4084.f13b}}
\parbox[e]{\textwidth}{\caption{{\bf a)} Profiles of the normalized \ha\
    surface brightness and residual velocity gradient parallel to the bar.
    The tracks of the profiles are the same as those in Fig.~\ref{perfilesA}a. {\bf b)} 
    Same as (a) but for the residual velocity gradient  perpendicular to the bar.
Note the offsets and anticorrelations of the
intense star forming regions and regions of high velocity gradient shown in
this figure and in Fig.~\ref{perfilesB}a (see text).}\label{perfilesB}}
\end{center}
\addtocounter{figmas}{1}
\end{figure*}

We saw qualitatively, by inspection in Sect.~\ref{sectiongrad}, that where the velocity
gradients in the bar are high, star formation appears less intense.
For a more quantitative appreciation of this effect we have made
graphs of residual velocity gradient and \ha\ intensity, taken along 
scans perpendicular to and parallel to the bar. We have used both Figs.~\ref{mosaico_barra}c 
and~\ref{mosaico_barra}d for this, so that we are comparing \ha\ surface brightness with 
velocity gradient parallel to the bar and perpendicular to the bar, respectively
in the two cross--sections in Figs.~\ref{perfilesB}a and  \ref{perfilesB}b.

In both graphs and in all the cross--sections, we note the coincidence
between the intensity peaks and local minima, or points of inflection, in the velocity gradient. 
There is a striking anticorrelation between velocity gradient and star
formation rate as represented by the \ha\ surface brightness, and it is particularly
striking in zones of high velocity gradient. In the two zones where the velocity gradient 
is both high and rapidly varying, between 10\arcsec\ and 15\arcsec\ from
the centre of the scans parallel to the bar 
through the nucleus (centre-right plot of 
Figs.~\ref{perfilesB}a and  \ref{perfilesB}b), 
there is a notable absence of star formation. These are the shocked zones at the edges of the central $x_2$
dominated region, where the flow along the bar is being braked and the gas is changing its 
orbital configuration. The star\- for\-ma\-tion within\- the central region is occurring
well within\- this radial range, i.e. well offset from these shocks. As well as
the global effects just described, there are local dips in the velocity
gradient, of smaller amplitude, coincident with the most intense star forming
re\-gions, which reflect the kinematic impact of the gas motions within the individual
regions on the overall velocity field.

\section{Velocity distribution and morphology within the inner disk}

\subsection{The nuclear bulge and the circumnuclear velocity pattern}

\begin{figure*}
\begin{center}
\resizebox{\textwidth}{!}{\includegraphics{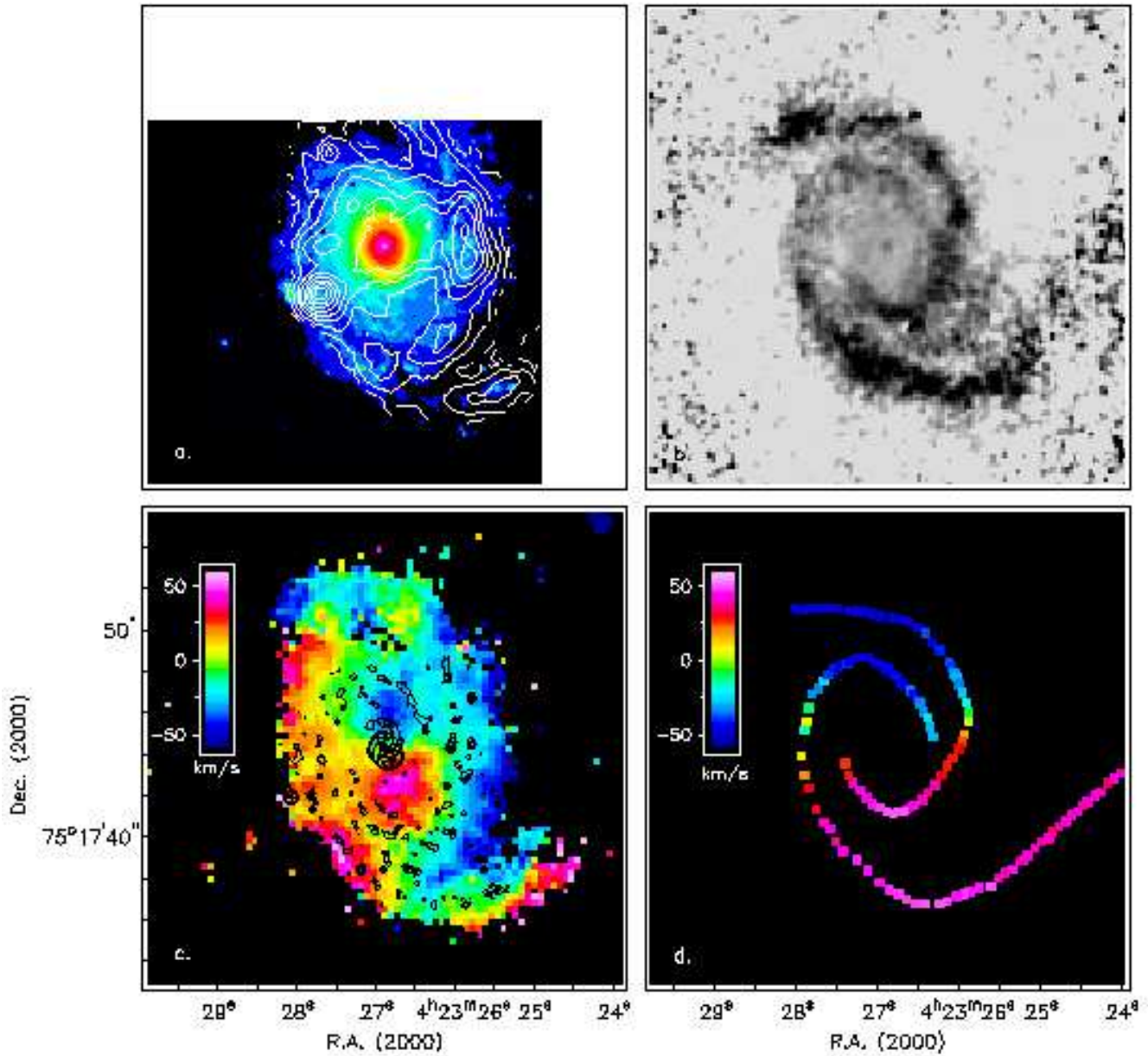}}
\parbox[e]{\textwidth}{\caption{{\bf a)} {\it HST}--NICMOS {\it H} band image of the central part 
of NGC~1530 with \ha\ isointensity contours overlaid 
at 1\arcsec\ resolution. {\bf b)} Grey--scale representations of the {\it J--K} colour index 
image. Redder colours are indicated by the darker shades. Image taken from fig.~2 of 
P\'erez--Ram\'\i rez et al. (2000). {\bf c)} Residual velocity map of 1\arcsec\ resolution 
of the central part of NGC~1530. This map was obtained after subtracting the model velocity field 
obtained from the modified rotation curve in the center from the velocity map (see text 
 details). Superposed are isointensity contours of the unsharp mask from the {\it HST}--NICMOS 
image shown in Fig.~\ref{unsharpmask}.
{\bf d)} Projected velocity field in the line of sight assuming a gas flow of constant amplitude 
(60~\kms) along the observed dust lanes (Fig.~\ref{mosaico_centro}b) in the inner part of 
NGC~1530. All images, a to d, show the same area of the centre of NGC~1530. }
\label{mosaico_centro}}
\end{center}
\addtocounter{figmas}{1}
\end{figure*}

Both morphologically and kinematically the central zone, 
which shows up as a projected disk of major axis radius $\sim10$\arcsec, i.e. $\sim1.8$~kpc, is 
easily distinguished in \ha\ from the bar. At its centre, as 
well seen in the Hubble Space Telescope ({\it HST})--NICMOS image presented in
Fig.~\ref{mosaico_centro}a, there appears to be a very 
small bulge, less than 2\arcsec\ in diameter, detectable via its circular isophotal 
shape. The rotation curve for the galaxy as a whole, which we presented in
Fig.~\ref{crmodified}, shows an apparently 
enhanced rotational velocity gradient within the inner $\sim$~5\arcsec, but we had produced this 
curve on the assumption that the only significant velocity component 
here is in the tangential direction to circular orbits in the plane of the galaxy. 

\begin{figure}
\begin{center}
\resizebox{8cm}{!}{\includegraphics{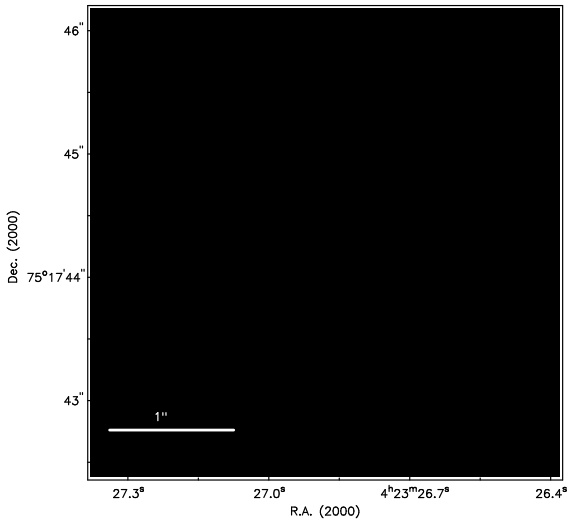}}
\parbox[e]{7.5cm}{\caption{Unsharp masked NICMOS image (Fig.~\ref{mosaico_centro}a)
of the central part of NGC~1530.}
\label{unsharpmask}}
\end{center}
\addtocounter{figmas}{1}
\end{figure}

From Fig.~\ref{mosaico_centro}a we find that there is no major extra mass component 
on the scale of this 
steeper gradient, so we explored the possibility that it is in fact indicating the presence 
of non--circular motions within this inner disk. To do this we assumed, from
the form of the rotation curve as a whole, that the velocity gradient between
5\arcsec\ and 10\arcsec\ represents the true rotation curve, and can be
in\-ter\-po\-lated between this portion of the curve and the nucleus. We then
subtracted off the rotated in\-ter\-po\-la\-ted curve, to produce a 
new residual velocity pattern for the central disk, which is shown in 
Fig.~\ref{mosaico_centro}c. We can see 
from this fi\-gu\-re that as well as the previously identified non--circular velocities beyond 
10\arcsec\ from the nucleus, we now have two lobes of oppositely directed velocity, nearly 
symme\-tric, to the north and south of 
the nucleus, with peak projected values of $\sim$50~\kms\ at a little 
under 5\arcsec\ from the centre. This pattern might be attributable to an
inner bar or oval distorsion within this inner disk. To check this out we
made an unsharp masked image (see Erwin \& Sparke 2002) from the NICMOS near IR image shown in 
Fig.~\ref{mosaico_centro}a, by subtracting off from the original a  version
smoothed with a gaussian of 5~pi\-xels half--width ($\sim0.38$\arcsec\ FWHM). The result 
is shown in  Fig.~\ref{unsharpmask}, in which we can see a bright nucleus, and 
a~\-small~\-asymmetric spiral, possibly e\-merging from a tiny bar, whose major axis appears
to be at PA$\sim120$\deg, but which is not well enough resolved even at this 
resolution.

The dynamical scope of this structure can be appreciated from
Fig.~\ref{mosaico_centro}c, where
it has been plotted onto the residual velocity map of the inner disk, and
where we can see that it is far too small to be causing the lobed velocity 
structure, which demands a different explanation. 

\subsection{The inner kinematic structure: a spiralling inflow pattern}
We can go on to interpret the inner residual velocity 
 structure as follows. The gas flow
along the bar and its\- de\-viation into circumnuclear orbits can be followed 
via\- the pre\-sence of dust lanes which indicate compressed and shocked gas.
These dust lanes in the central disk of NGC~1530 were imaged by 
P\'erez-Ram\'\i rez et al. (2000; their fig.~2b) in the near IR, at subarcsec resolution, and 
are especially well seen in their {\it J--K} colour image, which we reproduce here
in Fig~\ref{mosaico_centro}b. Two clear dust lanes can be seen to 
the north and south of the inner disk, coming inwards along the direction of the bar 
major axis, then curling around in orbits which take them
to the far side of the nucleus in both ca\-ses, where their trajectories have swung through 
270\deg, and terminating on opposite sides of the nucleus at some 
three arcsec from the centre, along a line with PA$\sim$120\deg.
The positions of the centres of our flow lobes in Fig.~\ref{mosaico_centro}c 
coincide with the inward curving dust lanes, so we attribute them to gas flow
coinciding with these lanes. The observed velocity structure of the lobes is
strongly affected by the fact that we observe the velocities along a single
line of sight, and so we detect only their projected components. Where the
velocity vector of the inflowing gas is perpendicular to the line of sight
the flow cannot be detected. Thus, the lobe to the north of the nucleus is in
fact reflecting the final stages of the inflow from the southern edge of the
bar, and vice versa. The two flows spiral around one a\-no\-ther in an
in\-ter\-locking pattern. 
As a test of this prediction we have produced a simple kinematic model showing 
how these flows would appear in projected velocity. We have a\-ssumed for 
simplicity that the intrinsic velocity vector has a constant amplitude, equal to 60\kms, and 
projected it into the line of sight to the observer, using the inclination
and position angle of the galaxy, and assuming that the dust lanes in the 
observations from P\'erez-Ram\'\i rez et al. (2000) denote the lines of flow, and that the
spi\-ral~arms~are trailing (this implies that the western part of the galaxy is closer to 
the observer than the eastern part). The
resultant pa\-ttern of pro\-jected velocities is shown in Fig.~\ref{mosaico_centro}d. Although 
the pattern is confined to the trajectory of the dust lanes, it agrees
remarkably well with the projected residual velocity pattern in Fig.\ref{mosaico_centro}c in
almost all respects. The only differences are seen towards the N and S of the
projected inner disc, where the model predicts higher residual velocities 
than those shown in Fig.~\ref{mosaico_centro}d. 
We explain this as follows: in the radial 
range from 10\arcsec\ to 20\arcsec\ our best rotation curve, shown in
Fig.~\ref{crmodified}, gives an apparent peak rising to some 300~\kms. It is probable that 
the peak,
and the curve in this range, show enhanced apparent rotational velocity because
we are not able to eliminate non--circular components in this range. Only 
the inner disc emits here, and opening up the free angle further reduces the 
signal to noise ratio to unacceptably low values. If we were able to make a
rational interpolation between the rising rotation curve within 10\arcsec\
radius, and the slowly falling outer curve beyond 25\arcsec, we might have 
attempted a more accurate separation of circular velocity and non--circular 
re\-si\-dual velocity in the outer part of the inner disc. However such an 
interpolation would be so far from linearity that it would be little more than
guesswork. Although the \hi\ distribution does not have the gaps of our \ha\
image, the angular resolution in the H~I rotation curve given in Regan en al.
(1996) is no better than 13\arcsec\ so cannot supply the information we need.
The pro\-jected model field in Fig.~\ref{mosaico_centro}d  gives such good agreement over most 
of the spiral flow lines that we feel secure in claiming that the discrepancies
to the north and south of the inner disc are most probably due to the 
effects of rotation curve subtraction described here.  

\subsection{A scenario for circumnuclear spirals}
A generic explanation for the presence of circumnuclear spirals was
given by Englmaier \& Shlosman (2000). They showed that in 
a galaxy where there is a clearly identified ILR, the 
response of gas to forcing by a bar can give rise, for an appropriate 
central mass distribution and range of sound speed, to a 
spiral pattern in the gas which can be present all the way down to within 
tens of parsec of the nucleus. They predicted that the arm/interarm contrast
should be low, as found in near IR observations of these structures by Laine
et al. (1999), and by P\'erez--Ram\'\i rez et al. (2000). The general framework 
for this kind of spiral structure does appear to be present in NGC~1530, 
since the form of the rotation curve has been shown by both Reynaud \& 
Downes (1997) and by Regan et al. (1996) to give rise to a predicted ILR.
Reynaud \& Downes (1997) predict an outer ILR at $\sim$~1.2~kpc from 
the nucleus, and an inner ILR some 100~pc from the nucleus. Here all we wish
to assert is the consistency of our results with this type of models. We do not
attempt to extend the work by any type of detailed modeling exercise.

\section{Discussion and concluding remarks}
\label{discusion}
\subsection{Three kinematic regimes}
NGC~1530 is a dramatically barred galaxy with very well 
defined morphological features, so it is not surprising that there is 
considerable recent work in the literature on its gas kinematics. However,
the use of an \ha\ emission line map at $\sim$1\arcsec\ resolution over the major
part of the disc has enabled us to probe more deeply some of its kinematical
properties. We have obtained a very well resolved rotation 
curve which is symmetric for the two independent halves of the galaxy. It was
easy to detect the presence of non--circular components of the internal motion,
in three dynamically distinct zones: streaming motions in the spiral arms, 
highly elliptical orbits around the bar, and an inwardly spiralling component
particularly marked in the circumnuclear disc within 2~kpc of the centre. By 
the conventional technique of subtracting off a two--dimensional projected model
rotation curve from the observed velocity field we produced a map of these
non--circular residual velocities, which include components of amplitude 
well over 100~\kms\ in the bar--induced elliptical orbits. The fact that the 
different kinematical regimes referred to above occur in distinct ranges of 
galactocentric radius, fa\-ci\-li\-ta\-ted the separation of the circular and the 
radial components. However both in the range of the spiral arms, from 63\arcsec\
to 90\arcsec\ from the centre, and in the inner disc, out to some 5\arcsec\ 
from the centre, we used auxiliary dynamical reasoning to effect an improved 
separation. We assumed that the departures from smooth radial dependence found
in these ranges cannot be due to rapidly radially varying mass distributions,
since no such variations are present in near--IR broad band images which trace the 
stellar mass, and must instead be due to projected components of local
non--circular motions not well separated out from the true rotation
curve. Based on this, we interpolated smooth underlying envelopes to the two
sections of the observed rotation curve, and used this smoothed version of
the rotation curve to produce revised maps of the residual velocity field, in which the streaming
motions in the arms and the spiralling motions towards the centre in the 
circumnuclear disc are better represented. These procedures do not affect the
strong residual velocities around the bar, which occur in an intermediate
range of radii, but do yield a more accurate overall map of the non-circular velocity
field.

\subsection{The key role of velocity gradients}
By far the most notable new results in this paper refer to the 
gradients in velocity which we have been able to detect directly and measure.
Particularly striking is the map of velocity gradients perpendicular to the 
bar, in the range of the bar (Fig.~\ref{mosaico_barra}d). Although in previous work on 
this and other barred galaxies the presence of a rapid jump in radial
velocities, from $\sim$+100~\kms\ to $\sim$-100~\kms\ as one transits the bar, has been 
detected, this is the first case where the relevant field has been closely 
mapped. It is essen\-tially a shear field, since when we compare this with the 
velocity gradient map in the direction along the bar, (i.e. along the direction 
of the gas flow vectors) the perpendicular gradient is considerably stronger.
We do note that a perfect separation into components parallel to and
perpendicular to the flow is not possible even in principle in observations 
purely of radial velocities, but that the dynamics of the situation, with the 
$x_1$ orbits around the outer lengths of the bar, is very favourable for a rather
good practical separation in NGC~1530. It is notable that the zones of
strongest gradient coincide precisely with the dust lanes even in fine
detail. 
%(our map has even better resolution than the optical broad--band image
%in Fig.~\ref{mosaico_barra}b where these lanes are well seen). 
 We have detected
gradients as steep as 0.35~\kms~pc\me\ in a direction which is essen\-tially
perpendicular to the flow vector along the bar, i.e. these are shear
gradients. Comparing this map with the \ha\ integrated intensity map in
Fig.~\ref{mosaico_barra}a we can see that there is a complete anticorrelation
between high values of shear gradient and the local star 
formation rate. Massive stars are not being formed in zones of high shear. This 
is not a surprising result at all. The work of Reynaud \& Downes (1998) 
was directed exactly at this point in NGC~1530, and came to this conclusion.

\subsection{Parallel shocks inhibit star formation; perpendicular shocks
  stimulate it}
The only advance we claim here is to have completed and detailed maps of the
phenomenology, with measured velocities and velocity gradients, both in
amplitude and direction. From these we can extract several interesting semi--quantitative 
conclusions. Firstly, although star formation region avoid strong shear, they 
do not necessarily avoid high non-circular velocity. Specifically the\-re is\- 
 a star\- for\-ming region, 20\arcsec\ to the NW of the galactic centre, in a region of
high velocity along the bar (greater than 100~\kms\ in a negative sense, deep 
blue in the residual velocity map). This region is, however, in a trough of 
velocity gradient: less than 0.05~\kms~pc\me\ both parallel to and perpendicular 
to the bar. It is interesting to note that this \hii\ region lies along the
trailing edge of the bar, and that this edge is the site of a number of
regions, while the leading edge of the bar does not show any strong
regions. This is not a common situation and could lead to a conclusion which
questions the placing of corotation near the end of the bar. However, we do
not believe that this is necessary. The bar is so strong that the spray
effect illustrated in Fig.~\ref{orbits}, should stimulate star formation along 
the trailing edge at positions of the arrow tips, where gas on
quasi-elliptical orbits collides with gas in the bar as it begins to reverse
its direction. Another way to consider this, is that we may be observing
effects of a 4:1 ultraharmonic resonance, but highly flattened by the strong bar.
All the other major star forming regions in the central disc, along
the bar, at the ends of the bar, and in the arms, occur in regions of low 
residual velocity as well as low velocity gradient. 

Secondly, there does appear
to be a re\-la\-tion\-ship between\- shocks (as picked out by steep velocity gradients
along the direction of the gas flow) and star forming zones. As seen particularly
well in the map of velocity gradients parallel to the bar (Fig.~\ref{mosaico_barra}c), linear
zones of high velocity gradient along the bar, at the edges of the central
disc, and even in the spiral arms (not shown in the cited figure) 
tend to accompany the regions of strongest star 
formation, with the zone of steep gradient offset from the centre of the nearby
star forming region by a distance of a few\- hundred\- par\-secs. This phenomenon could 
have two causes: either the compression along the line of flow is causing 
major gas condensation which leads to star formation, or outflow from star 
forming regions is impinging on the general flow to yield high gradients along
the flow direction. We have considerable evidence from detailed kinematical
stu\-dies of individual luminous \hii\ regions in a number of galaxies
(Rela\~no et al. 2003a, 2003b) that outflow at velocities of between 50~\kms\
and 100~\kms\ is occurring in most of them, so the second phe\-no\-me\-non could\- 
well occur. However, the scales of some of the gradients seen in
Fig.~\ref{mosaico_barra}c and the fact that in 
some cases these occur offset to one side of a star forming region and not 
to the other indicate that pre--stellar compression shocks are also being 
observed here. Quantification and modelling are clearly beyond the scope of 
the present paper, but these data may well offer a possible valuable route to
the spe\-ci\-fi\-ca\-tion of star formation conditions. It is, of course, the
difficulties of pinning down observationally the conditions prior to star
formation rather than any lack of theoretical offer which limits our
understanding of the phenomenon.

\subsection{Quantified spiral inflow towards the nucleus}
A two--dimensional map of the non--circular velocities also gives us 
the chance to follow kinematically the gas flow  which is picked out 
morphologically by the dust lanes. Although projection effects are quite 
severe on observations of a flow whose vector in the plane of the galaxy is
swinging through more than 270\deg, we have prima facie evidence that we
are able to follow this flow down to within 300~pc of the centre in this
galaxy. This is not very close to the nucleus, but bearing in mind that
NGC~1530 is 37~Mpc from us we could certainly expect interesting 
results from this method for closer objects.

\begin{acknowledgements}

We acknowledge Dr. M. Regan for kindly providing us his \ha\ rotation curve data,
Dr. D. P\'erez--Ram\'\i rez for help with observations 
%and permission to publish Fig.~\ref{mosaico_centro}b 
and Dr. S. Laine for directing our atten\-tion to the material in
Figs.~\ref{unsharpmask} and~\ref{mosaico_centro}a and for dis\-cussions on
its intepretation. 
We thank AURA for the use of the ima\-ge of NGC~1530 in Fig.~\ref{grad+V}b. 
The WHT is o\-pe\-ra\-ted on the island of La Palma by the Isaac Newton Group in
the Spanish Observatorio del Roque de los Muchachos of the Instituto de Astrof\'\i sica 
de Canarias.
Based on observations with the NASA/ESA Hubble
Space
Telescope, obtained from the data Archive at the Space Telescope Science
Institute, which is operated by the Association of Universities for
Research in Astronomy, Inc., under NASA contract NAS 5-26555.
This work was supported by the Spanish DGES (Direcci\'on General de 
Ense\~nanza Superior) via Grants PB91-0525, PB94-1107 and PB97-0219 and by
the Ministry of Science and Technology via Grant AYA2001-0435.

\end{acknowledgements}

\end{document}